\documentclass[twocolumn,showpacs,preprintnumbers,amsmath,amssymb]{revtex4}

\usepackage{graphicx}
\usepackage{dcolumn}
\usepackage{bm}
\begin{document}
\title{Wave function as geometric entity}
\author{B. I. Lev }

\affiliation{ Bogolyubov Institute for Theoretical Physics of the
NAS of Ukraine, 14-b, Metrolohichna Str., Kyiv 03680, Ukraine}
\date{\today}

\begin{abstract}
A new approach to the geometrization of the electron theory is
proposed. The particle wave function is represented by a
geometric entity, i.e., Clifford number, with the translation
rules possessing the structure of Dirac equation for any
manifold. A solution of this equation is obtained in terms of
geometric treatment. Interference of electrons whose wave
functions are represented by geometric entities is considered.
New experiments concerning the geometric nature of electrons are
proposed.

\end{abstract}
\maketitle

The problem of how to geometrize the electron theory and include
it in the scheme of the general relativity theory is far from
being solved. The expression for the covariant derivative
obtained in \cite{fok} from intuitional consideration, with some
interpretational corrections introduced in \cite{gel}, is the
generally accepted formula now. Cartan \cite{kar} has shown,
however, that finite-dimensional representation of a complete
linear group of coordinate transformations does not exist.
Moreover, the set of Dirac spinors preserves the structure of the
linear vector space, but does not preserve the ring structure
since defining the composition operation involves some
complications. Thus allowed states are depleted inasmuch as wave
function behavior under the parallel translation cannot be
calculated and appropriately interpreted, and, besides that, the
states of the particle ensemble cannot be determined.

A way to solve the problem is proposed in our paper. We employ
the idea of correspondence between the spinor matrices and the
elements of an exterior algebra and thus define the space of
states in terms of a space of representations of a space-time
Clifford algebra. Then the particle wave function is represented
by a complete geometric entity - a sum of probable direct forms
of an induced space of the Clifford algebra. This algebra
possesses ring structure \cite{kas} since it is a vector space
over the field of real numbers and hence makes an additive group
whose low of elements composition is distributive rather than
commutative with respect to addition. This ring has ideals which
may be obtained by multiplying the separated element on the right
or on the left by ring elements \cite{kas}. The ideals resulting
from this procedure are just the Dirac spinors of the standard
approach. Thus the representation of the Clifford algebra by the
Clifford number contains more information on particle properties
than spinor representations. Moreover, having attributed the wave
function with geometrical sense we can obtain correct translation
rules for an arbitrary manifold \cite{lev} and come to some new
quantum results associated with the geometric nature of the wave
function. Among these results we indicate the observation that
Dirac eqiuation in the geometric representation is nothing but
the translation equation in the general relativity sense, hence
its solutions may be interpreted geometrically. Moreover, the
geometric representation of the wave function yields other
results concerning the interference of elementary particles which
just may reveal the geometric nature of the wave function
\cite{kle}. We can also predict a new effect, i.e., time delay of
elementary-particle tunneling through a potential barrier which
can be explained only in terms of the approach proposed
\cite{las}.

We can introduce, at each point of the manifold, an arbitrary
basis that corresponds to the vector basis, and construct an
induced space of all possible products of basis vectors. Making
use of Clifford algebra for the composition of individual vectors
with simultaneously existing both inner and outer products, we
can set, at each point of the manifold, a unique complete
geometric characteristic - the direct sum of all possible
products for the elements of the induced space. The direct sum of
such tensor representations can be attributed with the Clifford
algebra structure by means of the direct product \cite{kas}. The
final dimensionality of the algebra is determined by the number
of basis vectors, provides the ring structure and is responsible
for the existence of an exact matrix representation. Moreover,
the space of functionals is isomorphic to this very linear space,
and the algebra of outer products is isomorphic to the algebra of
the outer product of these very vectors.

The existence of a unique set of linear independent forms defined
at an arbitrary point of the space suggests that the nature of
the forms translated over the manifold is similar to the nature
of forms which characterize it \cite{lev}. This may be also
determined by the similar form of the geometric entities as
functions of elements of the induced space. Suppose first we have
a basis determined by Dirac matrices $\gamma_{\mu}$. Making use
of this basis, we consider the realization of the wave function
in the ordinary Euclidean space. In this case the wave function
may be written in terms of a direct sum of a scalar, a vector, a
bivector, a trivector, and a pseudoscalar,
$\psi=\psi_{s}\oplus\psi_{v}\oplus\psi_{b}\oplus\psi_{t}\oplus\psi_{p}$,
that is given by
\begin{equation}
\psi=\psi_{0}\oplus\psi_{\mu}\gamma^{\mu}\oplus\psi_{\mu \nu
}\gamma^{\mu}\gamma^{\nu}\oplus\psi_{\mu\nu\lambda}\gamma^{\mu}\gamma^{\nu}\gamma^{\lambda}\oplus\psi_{\mu
\nu \lambda \rho
}\gamma^{\mu}\gamma^{\nu}\gamma^{\lambda}\gamma^{\rho}
\end{equation}
With the reverse order of composition, we have
$\tilde{\psi}=\psi_{s}\oplus\psi_{v}\ominus\psi_{b}\ominus\psi_{t}\oplus\psi_{p}$
and having changed the direction of each basis vector, we obtain
$\overline{\psi}=\psi_{s}\ominus\psi_{v}\oplus\psi_{b}\ominus\psi_{t}\oplus\psi_{p}$.
For each even number $\psi=\overline{\psi}$ for
$\psi\tilde{\psi}\neq 0$, the wave function may be reduced to the
canonical form \cite{kas}, i.e.,
\begin{equation}
\psi=\left\{\rho(x)\exp(i\beta)\right\}^{\frac{1}{2}}R
\end{equation}
where $R \tilde{R}=1$ describes all the coordinate
transformations associated with the translation and rotation of
coordinates and with the Lorentz transformation in the Euclidean
space. This Clifford number can be expressed in terms of an
exponential function of the biquaternion $B=q+iq'$, where $q$ and
$q'$ are quaternions, each of these representing a sum of a
scalar and a dual vector. The physical interpretation of this
geometric entity is rather evident since $\rho(x)$ can be
associated with the probability density of finding a particle in
an arbitrary spatial point, and $\beta$ is the angle that
determines the eigenstate of a particle with positive or negative
energy. We have $\beta=0$ for an electron and $\beta=\pi$ for a
positron. Thus it becomes possible to describe the intermediate
states of the particle since the form of the wave function of an
arbitrary ensemble of particles is analogous.

An arbitrary deformation of the coordinate system can be set in
terms of basis deformations $e_{\mu}=\gamma_{\mu}R$, where $R$ is
the Clifford number that describes arbitrary changes of the basis
(including arbitrary displacements and rotations) which do not
violate its normalization, i.e., provided $\tilde{R}R=1$. It is
not difficult to verify that
$e^{2}_{\mu}=\gamma_{\mu}\tilde{R}\gamma_{\mu}R=\gamma^{2}_{\mu}\tilde{R}R=I$
and this does not violate the normalization of the basis
\cite{kas}. Now, for an arbitrary basis, we can set, at each
point of the space, a unique complete linearly independent form
as a geometric entity that characterizes this point of the
manifold. For a four-dimensional space, such geometric entity may
be given by
\begin{equation}
\psi=\psi_{0}\oplus\psi_{\mu}e^{\mu}\oplus\psi_{\mu \nu
}e^{\mu}e^{\nu}\oplus\psi_{\mu\nu\lambda}e^{\mu}e^{\nu}e^{\lambda}\oplus\psi_{\mu
\nu \lambda \rho }e^{\mu}e^{\nu}e^{\lambda}e^{\rho}
\end{equation}
If this point of the manifold is occupied by an elementary
particle, then its geometric characteristics may be described by
the coefficients of this representation. A product of arbitrary
forms of this type is given by a similar form with new
coefficients, thus providing the ring structure. The operation of
form product may be written as
\begin{equation}
\psi \varphi=\psi\cdot\varphi+\psi\wedge\varphi
\end{equation}
where $\psi\cdot\varphi$ is an inner product or convolution that
decreases the number of basis vectors and $\psi\wedge\varphi$ is
outside product that increase number of basis vectors.

In order to determine the operation of form translation over an
arbitrary manifold we have to define the derivative operation. It
may written as a linear form $d=e_{\mu}\frac{\partial}{\partial
x_{\mu}}$ де $\frac{\partial}{\partial x_{\mu}}$ that forms a
basis of the vector space of all changes along the curves passing
through a given individual point of the space. The action of such
an operator on an arbitrary form may be presented as
\begin{equation}
d\psi=d\cdot\psi+d\wedge\psi
\end{equation}
where $d\cdot\psi$ and $d\wedge\psi$ may be called the
"divergence" and the "curl" of the relevant form.

The mapping of the manifold is determined by the mappings of the
relevant system of forms. A certain transformation group
transforms each form according to the law $\varphi'=\varphi R$,
where $R$ determines the mapping elements, of Clifford algebra in
our case, and satisfies the condition $\tilde{R}R=1$. Here action
of two successive transformations reduces to the action of the
third one, $RP=fQ$, in the Clifford algebra with appropriate
structure constants $f$. In this approach, all the elements of
the mapping and all the structure constants are expressed in
terms of Clifford numbers of general form with relevant tensor
characteristics. For this algebra, we can write the first
structure equation that defines the covariant derivative as given
by \cite{lev}:
\begin{equation}
\Omega=d\psi-\psi\omega
\end{equation}
with the gauge transformation law for the constraint $\omega$
being given by
\begin{equation}
\omega'=R\omega\tilde{R}+Rd\tilde{R}
\end{equation}
Here the tensor representation of the constraint is similar to
that of an arbitrary form of the Clifford algebra. In this case
the wave function can always be reduced to the canonical form,
but local deformations of the proper basis become, however,
inobservable since the Tetroude form $Rd\tilde{R}$ corresponds to
the second term of the gauge transformation. Then the second
structure equation that defines the "curvature" form may be
written as
\begin{equation}
F=d\omega-\omega\omega
\end{equation}
with the law of transformation under the algebra being given by
 $F'=R F\tilde{R}$.
This approach makes it possible to consider the mutual relation
of fields of different physical nature \cite{lev}. However, in
what follows we consider only the description of a particle wave
function as a geometric entity. The particle wave function is
described in terms of a geometric entity with the general
representation of the Clifford algebra. The group of possible
transformations of the frame of reference must transform the wave
function according to the structure equations given above. Under
the assumption that the covariant derivative $\Omega\sim m \psi$,
the first equation yields that the wave function should be
transformed according to the equation
\begin{equation}
d\psi-\psi\omega=m\psi
\end{equation}
whose form is analogous to the Dirac equation in the spinor
representation. The dynamic equation for wave function in
geometrical presentation can obtain from action which can present
in the term of geometrical invariants as follow:
\begin{equation}
S=\int d\tau\left\{\Omega
\widetilde{\Omega}+F\widetilde{F}\right\}
\end{equation}
using the normalization condition $\int d\tau
\psi\widetilde{\psi}=1$,where $\tau$ is volume of four
dimensional space.

This equation is more informative for several reasons. The first
one is that spinors are only special projections of Clifford
numbers \cite{kas}, Dirac spinors are represented only by ideals
in this algebra, and thus it is impossible to introduce the
composition operation on the spinor set. And the most important
difference is that complete group of linear transformations of
the coordinate system does not exist for spinors \cite{kar}. As
follows from the previous analysis, a complete transformation
group associated with the structure equation exists only in the
Clifford-number representation of the wave function. The first
structure equation for the wave function reproduces the form of
the Dirac equation and, as it has been shown in \cite{kas}, its
solutions are similar to those for the spinor representation,
This solves the problem of finite-dimensional representation of
the wave function under the complete linear group of coordinate
transformations.

For illustration this approach can present the solution of Dirac
equation  in geometrical presentation. Can consider the behaviour
the electron in Coulomb field for relativistic case. In
time-space of special relativity theory
$d=\gamma_{\mu}\frac{\partial}{\partial x_{\mu}}$ a
$\omega_{\mu}\equiv A_{\mu}$ where $
A_{\mu}=-\gamma_{0}\frac{Z\alpha}{r}=\gamma_{0}U$ is the vector
potential,where $\alpha$ is constant of subtle structure and
$m=i\mu \gamma_{0}$ where $\mu $ is mass of electron. The
structure equation thus obtained is written in the introduced
terms is completely equivalent to the Dirac equation, and has
well known solutions both for the calculation of the hydrogen
atom spectrum and for the interpretation of electron states
\cite{kas}. Multiplied the structural equation at the left on
$\gamma_{0}$. After multiplication operator $d$ transform to
$\gamma_{0}d=\frac{\partial}{\partial x_{0}}+\mathbf{\nabla}$
where spatial gradient
$\mathbf{\nabla}=\sigma_{\mu}\frac{\partial}{\partial x_{\mu}}$
where $\sigma_{\mu}$ is Pauli matrix. The solution of new equation
can present in the form $\psi=\left(q+q'\right)\exp(iEx_{0})$
where $q$ and $q'$ is quaternions, $E$ is energy  and $x_{0}=ct$
is time coordinate. After this the Dirac equation reduce to two
intercoupling equations:
\begin{equation}
\nabla q=\left(\mu+E-U\right)q'
\end{equation}
\begin{equation}
\nabla q'=\left(\mu-E+U\right)q
\end{equation}
The solution of this system equation can be writhe in the form
$q=iSG(r)$ and $q'=iPF(r)$ where $G(r)$and $F(r)$ is number
functions and $S=\sigma_{r}P(\theta,\varphi)$ where
$P(\theta,\varphi)$ quaternion which dependence from polar and
azimuthal angle in spherical coordinate. Substituted this
presentation in system of equation for quaternions can obtain the
specter of energy in standard form:
\begin{equation}
E=\mu\left\{1+\frac{(Z\alpha)^{2}}{\sqrt{l^{2}-(Z\alpha)^{2}}+n^{2}_{r}}\right\}^{-\frac{1}{2}}
\end{equation}
where $n_{r}$ radial quantum number and $l$ orbital quantum
number. Obtained result confirmed the consistency between
standard and present approach. The new thing here is that this
equation can be solved for a system with both electric and
gravitational fields. We shall not do that since the
gravitational field only weakly influences the atomic state and
it is very difficult to find the evidence for the geometric
nature of the electron. If used the canonical presentation of
wave function in the form (2) and assume that $g_{\mu
\nu}=\rho(x)G_{\mu\nu}$ where $G_{\mu \nu}$ is potential of
repairable gravitational field from action (10) can obtain the
system of covariation equations in the form:
\begin{equation}
\theta_{\mu \nu}-\frac{1}{2}\theta g_{\mu
\nu}=\frac{1}{6}\left(R_{\mu \nu}-\frac{1}{2}R g_{\mu \nu}\right)
\end{equation}
and
\begin{equation}
\frac{\partial}{\partial x^{\mu}}\left(\frac{\partial
\beta}{\partial x_{\mu}}\right)
\end{equation}
where $\theta_{\mu \nu}=\frac{\partial \beta}{\partial
x_{\mu}}\frac{\partial \beta}{\partial
x_{\nu}}-\frac{1}{4}\mu^{2}G_{\mu \nu}$, $\theta=\theta_{\mu
\nu}g^{\mu \nu}$ , $R_{\mu \nu}$ is curvature tensor.This system
is whited in presentation $\hbar=c=1$. For
$\beta=\frac{S}{\hbar}$ as standard approach obtained system
equations describe the intercoupling between energy-impulse of
moving particle with metrical tensor produced it the time-space.
Indeed, that since the gravitational field only weakly influences
the electron state and it is very difficult to find the evidence
for the geometric nature of the electron. Instead we consider a
simpler effect, i.e., interference of electrons or other
elementary particles whose geometric nature can be revealed by
available experimental methods.

We consider the case of particle interference that might be
helpful in revealing the geometric character of the wave
function. In our representation, two elementary particles can be
described by the wave functions represented by geometric entities
in the canonical form, i.e,
$\psi_{1}=\left\{\rho_{1}(x)\exp(i\beta_{1})\right\}^{\frac{1}{2}}R_{1}$
and
$\psi_{2}=\left\{\rho_{2}(x)\exp(i\beta_{2})\right\}^{\frac{1}{2}}R_{2}$.
For electrons we have $\beta_{1}=\beta_{2}=0$. The canonical form
of the two-electron wave function should be similar, i.e.,
$\psi=\left\{\rho(x)\exp(i\beta)\right\}^{\frac{1}{2}}R=\psi_{1}+\psi_{2}$.
Now the post-interference wave function can be written as
\begin{equation}
\psi\tilde{\psi}=\rho(x)=\rho_{1}+\rho_{2}+\left(\rho_{1}\rho_{2}\right)^{\frac{1}{2}}\left\{R_{1}\tilde{R_{2}}+\tilde{R_{1}}R_{2}\right\}
\end{equation}
In the case of even Clifford numbers, when $R$ corresponds to
Lorentz rotations, i.e., when $R_{i}$ can be written as
$R_{i}=\exp(-B)$, where $B=(\theta+i\varphi)b$ is a double
vector, $\theta$ and $\varphi$ are constant numbers, and $b$ is a
vector whose modulus is equal to one, the result of interference,
for equal particle energies, is given by the standard expression,
i.e.,
\begin{equation}
\psi\tilde{\psi}=\rho(x)=\rho_{1}+\rho_{2}+\left(\rho_{1}\rho_{2}\right)^{\frac{1}{2}}\cos\varphi
\end{equation}
For plane monochromatic waves \cite{kas}, the solution of the
Dirac equation is given by
 $\psi_{1}={\rho_{1}(x)}^{\frac{1}{2}}u
\exp(i\sigma_{3}(p \cdot x))$, where $\sigma_{3}$ is the Pauli
matrix, $u$ is particle amplitude, and $p$ is particle momentum.
The solution for the second particle is similar except for the
phase shift, i.e., we have
 $\psi_{2}=
{\rho_{2}(x)}^{\frac{1}{2}}u \exp(i\sigma_{3}(p\cdot
x)+\varphi)$. We see that now electron interference is described
by the well known formula. Next we assume that deformations of
the reference system are determined by both even and odd numbers
and that each transformation contains both even and odd parts,
i.e., $R_{i}=R^{ev}_{i}+R^{od}_{i}$, which could be produced by
fields of different nature whose effect on different geometric
components of the general Clifford number is different
\cite{lev}. Then the operation
\begin{equation}
R_{1}\tilde{R_{2}}+\tilde{R_{1}}R_{2}=R^{ev}_{1}R^{ev}_{2}-R^{od}_{1}R^{od}_{2}
\end{equation}
reduces to two terms and we see that the result of electron
interference is described by an essentially nonstandard formula.
The second term in the right-hand part of the equation also
reduces to an even Clifford number, but possesses different
structure.

The existence of this effect can be verified experimentally. A
coherent electron beam should be divided into two beams, the
latter should be passed through separate regions with variable
basic geometric characteristics. The change of the wave function
passing through different regions can be written as
 $\psi'_{i}=\psi_{i}R_{a}R_{b}$ where
$R_{a}$ and $R_{b}$ describe the transformation of particle
characteristics in the regions $a$ and $b$. If the sequence order
is changed, $\psi''_{i}=\psi_{i}R_{b}R_{a}$ $R_{a}R_{b}\neq
R_{b}R_{a}$ then electron interference should correspond to the
last case of the previous analysis, i.e., the interference
pattern should be different from the standard case. The various
regions can be infinite solenoids of the Aaronov-Bohm experiment
with different directions of the magnetic flux. Another way to
observe the difference of the interference patterns is to pass
electrons along and across the solenoids. The difference is given
rise to only by the geometric representation since in the first
case the flux is not changed as distinct to the opposite case. We
can assume that this effect might be also observed for neutron
interference, the regions of variation of the wave function
geometric components being two inclusions with different mass
numbers occurring on the neutron propagation path. A similar
experiment had been proposed in paper \cite{kle}, however, it has
not been performed till now.


\begin{thebibliography}{99}

\bibitem {fok} W. A. Fock, Zs.for Physics \textbf{57}, 2611, (1929).

\bibitem{gel}  V. A. Zhelnorovich,{\it Theory of spinors and application in mechanics and physics},
(Nauka, Moskow, 1982).

\bibitem{kar}  E. Cartan {\it Lecons sur ia theorie des spineurs}(Actualites scientifiques et industries, Paris, 1938).

\bibitem{Sch}  B. F. Schutz {\it Geometrical mathods of mathematical physics}(Cambridge University Press, Cambridge, 1982).

\bibitem {kas} G.Kasanova, Vector algebr, (Cambridge University Press, Cambridge,(1979)

\bibitem {tak} J. M. Benn and R. W. Tucker, Phys. Lett.A \textbf{130},177 (1983).

\bibitem {lev} B. I. Lev, Mod. Phys. Lett. \textbf{3},10, 1025,(1988),ib.\textbf{4}, eratum, (1989)

\bibitem {kle} A. G. Klein, Physics B, \textbf{151}, 44, (1988).

\bibitem {las} J. Lasenby, A. N. Lasenby and J. I. Dosan, Phil Trans. R. Soc.Lond., 1, (1996)
\end{thebibliography}
\end{document}